\documentstyle[aps,twocolumn,prb,epsf]{revtex} 

\begin{document}  \preprint{\today} \draft
\title{Spin and orbital excitations in magnetic insulators with Jahn-Teller ions}

\author{J. van den Brink, W. Stekelenburg, D. I. Khomskii 
and G. A. Sawatzky}

\address{
Laboratory of Applied and Solid State Physics, Materials Science Centre,\\
University of Groningen,Nijenborgh 4, 9747 AG Groningen, The Netherlands}

\author{K. I. Kugel}
\address{
Scientific Physics Institute,\\ 
Russian Acad. Sci., Lininskii pr. 53, Moscow, 117924 Russia}

\maketitle

\begin{abstract}

The elementary excitations of a model Hamiltonian that captures the 
low energy behavior of a simple two-fold
degenerate Hubbard Hamiltonian with Hund's rule coupling, is studied. 
The phase diagram in the mean-field limit and in a two-site approach reveals a rich variety 
of phases where both the orbital and the spin degrees of freedom are ordered.
We show that besides usual spin waves (magnons) there exist also 
orbital waves (orbitons) and, most interestingly, 
in a completely ferromagnetically coupled system, a combined
spin-orbital excitation which can be visualized as a bound state of magnons and orbitons.
For a completely degenerate system the bound states are found to be the lowest lying 
elementary excitations, both in one- and two-dimensions.
Finally we extend our treatment to 
almost-degenerate systems.  This can serve as an example that the elementary excitations
in orbitally degenerate strongly correlated electron systems in general carry both spin 
and orbital character.

\end{abstract}


\section{Introduction}

The $d$ and $f$ wavefunctions of free atoms or ions are, besides Kramers (spin) degenerate, 
also 5 and 7-fold orbital degenerate.
In crystals this degeneracy may be lifted by the crystal field
(interactions with the ligands).
There are, however, interesting situations where this degeneracy is only 
partially lifted, so that in a high symmetry situation $d$- or $f$-levels remain degenerate.
The best known examples are Cu$^{2+}$ (d$^9$), Mn$^{3+}$ (d$^4$) and
Cr$^{2+}$ (d$^4$) in cubic symmetry (in octahedral surrounding).
According to the famous Jahn-Teller theorem~\cite{Jahn37} this degeneracy should be
lifted in the ground state. In a concentrated system this leads to the 
phenomenon known as the cooperative Jahn-Teller effect, or 
orbital ordering~\cite{Gehring75,KK78}. An interesting aspect of this phenomenon
is the strong interplay between the orbital and spin (magnetic)
ordering. The orbital occupation determines the
character of the magnetic exchange interaction 
(the Goodenough-Kanamori-Anderson rules)~\cite{Gooden63}; and, vice versa,
modification of the magnetic structure, e.g. by an external field, may change
orbital occupation and lead to a change in the crystal structure~\cite{KK78,KK80}.

The existence of orbital degrees of freedom, strongly interacting with the spins,
does not only determine the orbital and spin structure in the ground state.
It should also have important consequences for the elementary excitations of 
such systems. Thus, in addition to the collective excitations of the magnetic
subsystem -magnons, or spin waves-, also orbital excitations -orbitons- may 
exist in this case.
This was pointed out in ref.~\cite{KK78} and was studied for a specific model
for the manganites in ref.~\cite{Ishihara96}.
Because of the intimate connection of spin and orbital degrees of freedom,
one can also expect strong interaction and possible mixing of
these two types of excitations. This problem was not addressed until now,
and it is one of the aims of the present study.

The systems with orbital degeneracy form a rather special class of compounds
with very interesting and rich properties. Among them are, in particular, many
compounds containing Cu$^{2+}$ (like actually the high T$_{\rm c}$ superconductors),
manganites of the type La$_{1-x}$Sr$_{x}$MnO$_3$ with a colossal magneto-resistance,
but also many other transition metal compounds containing 
vanadium (LiVO$_2$, CaV$_4$O$_{9}$)~\cite{Pen97,Marini_tbp2},
trivalent nickel (PrNiO$_3$)~\cite{Garcia92} etc. The study of the
elementary excitations in these compounds, besides of being interesting
on itself, may shed some light on their unusual properties.

In this work we consider the typical situation of materials containing localized
electrons with spin S$=\frac{1}{2}$ with doubly degenerate orbitals.
We can describe this situation by the doubly degenerate Hubbard model:
\begin{eqnarray}
H = H_t + H_U +H_J
\label{eq:ham_general}
\end{eqnarray}
with
\begin{eqnarray}
&H&_t = \sum_{<ij>,\sigma,\alpha,\beta} t^{\alpha \beta}_{ij} \ c^{\tiny \dag}_{i \alpha \sigma} c_{j \beta \sigma} \\
&H&_U = \sum_{i,\sigma,\sigma',\alpha,\beta} U^{\alpha \beta} n_{i \alpha \sigma} n_{i \beta \sigma'} 
       (1-\delta_{\alpha,\beta} \delta_{\sigma,\sigma'}) \\
&H&_J = \sum_{i,\alpha,\beta} -J_H^{\alpha \beta} {\bf S}_{i \alpha} \cdot {\bf S}_{i \beta} (1-\delta_{\alpha,\beta}),
\end{eqnarray}
in which, besides the usual terms (electron hopping and on-site Coulomb repulsion),
we also added the on-site (Hund) exchange interaction.

In realistic situations the hopping matrix elements $t^{\alpha \beta}_{ij}$ depend
on the type of orbital involved~\cite{KK78,Gooden63,KK80,Ishihara96}.
This leads to enormous technical complications. As here we are interested in
the basic typical features of the excitation spectrum of our system, we 
treat a simplified symmetric model, assuming 
\begin{eqnarray}
t^{\alpha \beta}_{ij} = t_{ij} \delta_{\alpha \beta},
\label{eq:hop}
\end{eqnarray}
keeping only the nearest neighbor hopping for equal orbitals. At the end of the
paper we  comment which modifications of our results should occur for 
a more general choice of hopping integrals.
In the case of strong interaction $t \ll U$ only the degrees of freedom of
the localized electrons with spin $\frac{1}{2}$ and the orbitals are left.
Consequently we can reduce the electronic model~(\ref{eq:ham_general}) to a
model describing coupled spins and orbitals. For the doubly degenerate case we can
describe the orbital degrees of freedom by an effective pseudo-spin T$=\frac{1}{2}$,
so that e.g. an occupied orbital $\alpha$ corresponds to T$^{\rm z}=\frac{1}{2}$ and
orbital $\beta$ to T$^{\rm z}=-\frac{1}{2}$. The effective Hamiltonian has the 
generic form
\begin{eqnarray}
H &=& -J_s \sum_{<ij>} {\bf S}_i \cdot {\bf S}_j  
      -J_t \sum_{<ij>} {\bf T}_i \cdot {\bf T}_j  \nonumber \\
  &-& 
       4J_{st} \sum_{<ij>} ( {\bf S}_i \cdot {\bf S}_j ) \ ( {\bf T}_i 
       \cdot {\bf T}_j ),
\label{eq:HST}
\end{eqnarray}
where for the model with $t^{\alpha \beta}_{ij}$ given by~(\ref{eq:hop})
the exchange constants have definite values~\cite{KK78}. Thus for this
particular choice of hopping matrix elements the model~(\ref{eq:ham_general})
reduces to a double-spin Heisenberg model.

The characteristic feature
of our situation is the existence and strong interplay of 'spins' $S$ and $T$
which emerges due to the third term of Hamiltonian~(\ref{eq:ham_general}).
As we shall see, this interaction leads to a significant mixing of
spin and orbital interaction, giving rise, even, to the possibility
of the formation of spin-orbiton bound states. 
Note that the form of the interaction between the spins and orbitals 
in~(\ref{eq:ham_general}) is a consequence of the Coulomb and Hunds rule
interaction between electrons and is different from 
what one expects from simply coupling two Heisenberg spin systems.
As we want to study the generic
features of the coupled spin-orbital system, we consider the general case,
treating arbitrary values of the exchange parameters $J_s$, $J_t$ and $J_{st}$.
Notice that in the general situation with Hund's rule exchange included and
with realistic values of the hopping integrals $t^{\alpha \beta}_{ij}$ the resulting
spin-orbital Hamiltonian of the type~(\ref{eq:HST}) can contain terms anisotropic
in the $T$-operators~\cite{KK78} and even terms linear in $T$. However, even
with the more realistic hopping integrals there may exist situations~\cite{Feiner97} in which 
the symmetry in orbital space remains similar to that of Hamiltonian~(\ref{eq:HST}). 
For simplicity we consider such a symmetric double-Heisenberg model because,
as it turns out, the conditions for the existence of the bound spin-orbital
excitations are the most stringent exactly in this case, and if, as we will show,
they exist in this case, it will be even more so with the $T$-anisotropy taken
into account.

\section{Ground state}

If our aim is to calculate the elementary excitations of a system described by the
Hamiltonian~(\ref{eq:HST}), we first have to know the ground state wavefunctions.
We can easily calculate a phase diagram  within two different approximations: the mean-field
and the two-site approximation. In the latter only two interacting spins and pseudo-spins
are considered, so that the ground state is described in a way similar to a valence-bond
state, which is a good starting point for low-dimensional spin systems. Starting from
ground state wavefunctions obtained within one of these approximations, one can test the
stability of the ground state by calculating its response to for instance spin waves and
combined orbital- and spin-excitations.

\subsection{Mean Field Solution}

A common method used to gain some understanding of a Hamiltonian is  a mean field
approximation. In our case it is possible to separate spin and orbital degrees of
freedom  by replacing operators we want to exclude from our consideration by 
the averages of their correlation function that appears in the Hamiltonian.
In this way we generate two mean-field Hamiltonians
\begin{eqnarray}
H^{MF}_s &=& -J_s \sum_{<ij>} {\bf S}_i \cdot {\bf S}_j  
  -4J_{st} \sum_{<ij>} <{\bf T}_i \cdot {\bf T}_j>   {\bf S}_i \cdot {\bf S}_j \nonumber \\
  &=& -J'_s \sum_{<ij>} {\bf S}_i \cdot {\bf S}_j
\label{eq:HS_mf}
\end{eqnarray}
and
\begin{eqnarray}
H^{MF}_t &=& -J_t \sum_{<ij>} {\bf T}_i \cdot {\bf T}_j  
  -4J_{st} \sum_{<ij>} <{\bf S}_i \cdot {\bf S}_j>  {\bf T}_i \cdot {\bf T}_j \nonumber \\
  &=& -J'_t \sum_{<ij>} {\bf T}_i \cdot {\bf T}_j.
\label{eq:HT_mf}
\end{eqnarray}

\begin{figure}
      \epsfysize=40mm
      \centerline{\epsffile{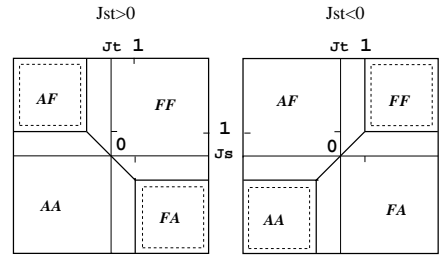}}
\caption{Phase diagram of the model Hamiltonian in the mean field approximation. FF: both spin
and pseudo-spin are ferromagnetically ordered; FA: spin ferro, pseudo-spin anti-ferro; AF: spin anti-ferro,
pseudo-spin ferro; AA: both spin and pseudo-spin are anti-ferro ordered. $J_s$ and $J_t$ are in
units of $|J_{st}|$. }
\label{fig:phases_MF}
\end{figure}

In Hamiltonian~(\ref{eq:HS_mf}) the orbital degrees of freedom are integrated 
out and in 
Hamiltonian~(\ref{eq:HT_mf}) the average over the spin degrees of freedom is taken.
In this way the spin and orbital degrees of freedom are decoupled.
After doing the mean-field averaging~\cite{Ishihara96} we have, 
in some sense, thrown away the interesting part 
of our problem and returned to a renormalized Heisenberg model, where the ground state properties
and elementary excitations are well known. Nevertheless, as in a first approximation it
is still a useful approach. The phase diagram in this approximation is given in figure~\ref{fig:phases_MF}.

\subsection{Two-Site Solution}

\begin{figure}
      \epsfysize=40mm
      \centerline{\epsffile{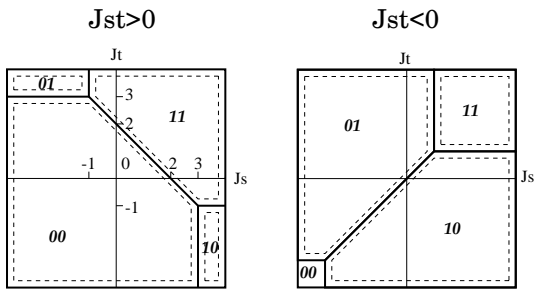}}
\caption{Phase diagram of the two-site solution of the model Hamiltonian. In the 11 phase both spin
and pseudo-spin are in a triplet state, in the phase 10 the spin is triplet, pseudo-spin singlet,
in the phase 01 the spin is singlet, pseudo-spin triplet and in the phase 00 both spin and pseudo-spin are
in a singlet state. $J_s$ and $J_t$ are in units of $|J_{st}|$.}
\label{fig:phases_2p}
\end{figure}

\begin{figure}
      \epsfysize=40mm
      \centerline{\epsffile{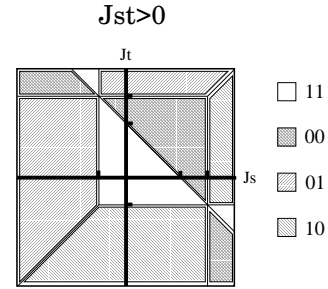}}
\caption{Character of the first excited state in the two-site system. 
$J_s$ and $J_t$ are in units of $|J_{st}|$.}
\label{fig:exc_2p}
\end{figure}

In a mean-field approximation the short range interactions are averaged out, and if we want to 
go beyond this approximation and gain insight into local properties it is more useful to 
consider a few interacting particles. This is especially important here because the transformation
properties of these terms of the Hamiltonian~(\ref{eq:HST}) are different, and the more accurate
account of the real singlet correlations (as compared to the mean field, essentially Ising-like, treatment)
is essential.
To this end case we consider two spins and
two pseudo-spins and obtain the Hamiltonian
\begin{eqnarray}
H_{1,2} = &-&J_s {\bf S}_1 \cdot {\bf S}_2  
      -J_t  {\bf T}_1 \cdot {\bf T}_2  \nonumber \\
      &-&4J_{st}  ( {\bf S}_1 \cdot {\bf S}_2 ) \ ( {\bf T}_1 \cdot {\bf T}_2 ).
\label{eq:H_12}
\end{eqnarray}
This simple case can be treated exactly, and using these results we obtain the modified
phase diagram. The combinations of (pseudo-)spin operators can either be singlet or triplet and the
ground state energy of the various combinations correspond to the different phases in figure~\ref{fig:phases_2p}, 
for $J_{st} > 0$ and $J_{st} < 0$, respectively.
The character of the first excited state for the various phases for $J_{st} > 0$ is  shown in 
figure~\ref{fig:exc_2p}. A similar picture can be made for $J_{st} < 0$.
In spite of the relative simplicity of the Hamiltonian, there is a rich 
variety of ground- and first-excited states, even in the two-site approach.

In the white, central part of this figure the ground state is singlet,
both in the spins and orbitals. The first excited state, however, is
not just the change of one singlet into a triplet, but corresponds to the state
where both the spins and orbitals are in a triplet configuration. This is due
to the interaction between the spin and orbital degrees of freedom and
this very simple example shows that some of the excitations of a system with such
an interaction, in this case the lowest one, carry both spin and orbital character.

\section{Ferro-Ferro system}
When $J_s$, $J_t$ and $J_{st}$ are chosen in a range where both the spins and pseudo-spins order ferromagnetically,
a ferro-ferro phase, we can obtain exact analytical expressions for the elementary excitations. In a ferromagnetic
Heisenberg model the exact ground state is the state with all spins pointing in the same direction, as opposed to
a Heisenberg Hamiltonian with an antiferromagnetic exchange where quantum fluctuations affect the Ne\'el spin order.

\begin{figure}
      \epsfysize=40mm
      \centerline{\epsffile{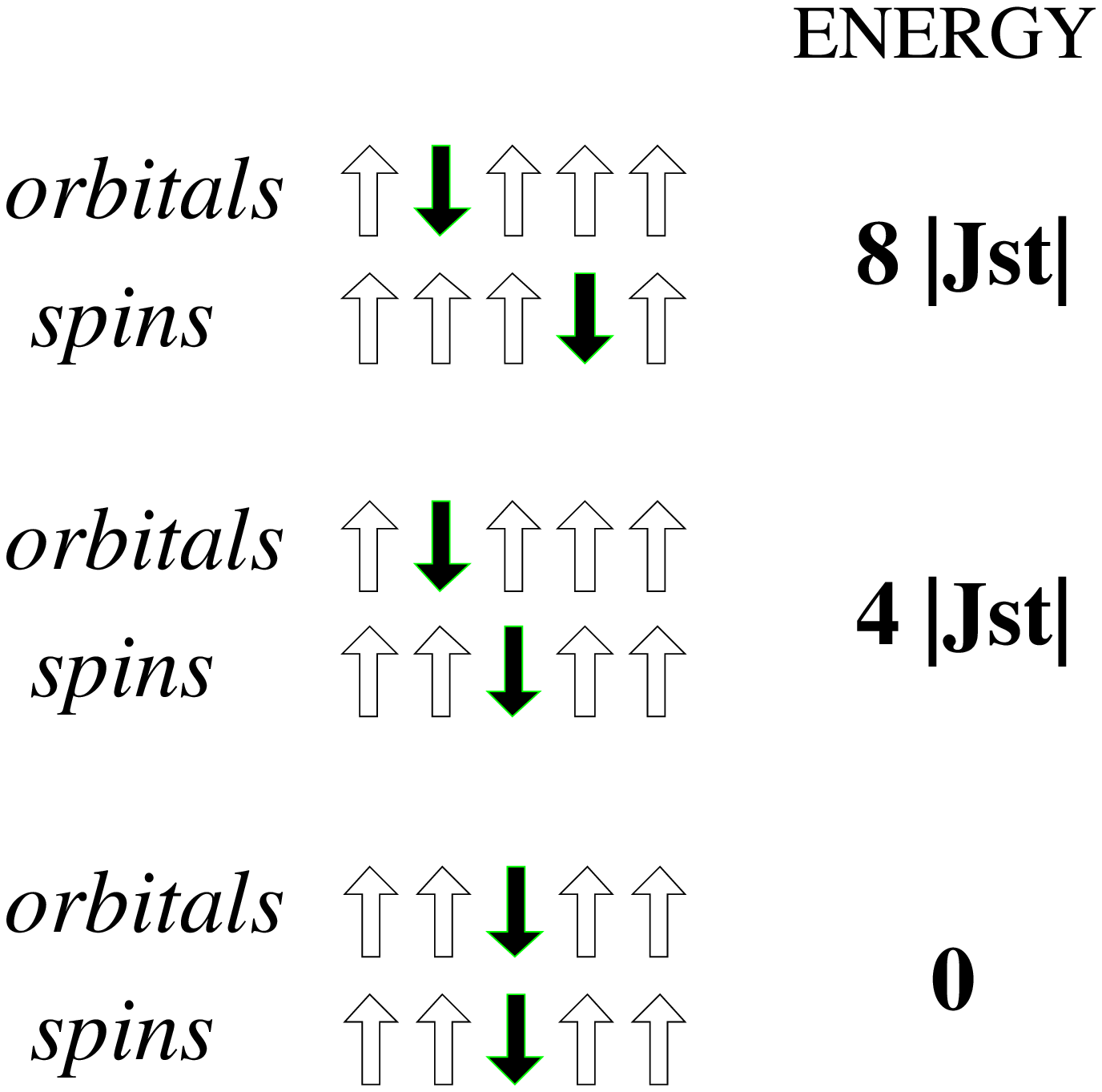}}
\caption{Excitation energy for Ising spins and pseudo-spins for the ferro-ferro
system. When the spin and orbital are close together, the excitation energy is lower
than in the case they are far apart.}
\label{fig:sp_orb}
\end{figure}

The coupling between the spins and pseudo-spins can lead to the formation of bound states 
of spin and pseudo-spin excitations. This is illustrated for the ferro-ferro
system in figure~\ref{fig:sp_orb}. Let us consider Ising spins (and pseudo-spins) and examine the
energy of an excitation of one spin plus one pseudo-spin. If the two excited spins are
far away from each other, the part of the excitation energy due to 
the spin pseudo-spin interaction (the third term in~(\ref{eq:HST}))
is $8 |J_{st}|$.  When the excited spins are close to each other, on the nearest neighbor sites
or on the same site, this excitation energy is $4 |J_{st}|$ and $0$, respectively. 

This means that a magnon and an orbiton (orbital wave) have an attractive interaction and that
in principle, depending on dimensionality and strength of the attraction, the magnon and orbiton can form
a bound state.

\subsection{Equations of motion}
In order to establish whether it is indeed possible to obtain bound states that 
are combinations of orbital- and spin waves
in the excitation spectrum of the model system, we  use the equation of motion method, which provides
a simple, and in the ferromagnetic case exact, way to calculate bound state energies. 
Before addressing this issue, let us first examine a single (pseudo-)spin excitations.

Starting from Hamiltonian~(\ref{eq:HST}) and with a ground state $|0>$ with 
energy $E_0$ where all (pseudo-)spins are aligned in the 
positive $z$ direction, a single spin is excited (the derivation for single orbital excitations is equivalent).
The equation of motion for this excitation is:
\begin{eqnarray}
(H-E_o) S^-_m \ |0> \equiv \omega_s S^-_m  \ |0>  = [H,S^-_m] \ |0>.
\label{eq:EQM_spin}
\end{eqnarray}
The evaluation of the commutator and transformation to Fourier-space yields the following Goldstone
modes:
\begin{eqnarray}
\omega_s({\bf Q}) &=& 2 (J_s+J_{st})  \sum_{\bf a} ( 1-\cos {\bf Q \cdot a} ) \\
\omega_t({\bf Q}) &=& 2 (J_t+J_{st})  \sum_{\bf a} ( 1-\cos {\bf Q \cdot a} ),
\label{eq:om_orb}
\end{eqnarray}
where the lattice vectors are denoted by ${\bf a}$. The presence of the coupling between the spin and orbital
degree of freedom, parameterized by $J_{st}$, thus merely renormalizes the spin/orbital wave spectrum.

The equation of motion for a combined spin and pseudo-spin excitation $S^-_m T^-_n$ is
\begin{eqnarray}
(H-E_o) S^-_m T^-_n \ |0>       &\equiv& \omega S^-_m T^-_n \ |0>  \nonumber \\ 
                                &=& [H,S^-_m T^-_n] \ |0>.
\label{eq:EQM}
\end{eqnarray}
We find for the Fourier transform of the equation of motion 
\begin{eqnarray}
   \{\omega &-& \gamma({\bf Q,q} \} ) A({\bf Q,q}) = \nonumber \\
    &-&8 J_{st} \ (\frac{a}{2 \pi})^d \sum_{\bf a} 
                (\cos {\bf Q \cdot a} /2 - \cos {\bf q \cdot a } ) \nonumber \\
        &\cdot& \int d{\bf k}( \cos {\bf Q \cdot a} /2 -\cos {\bf k \cdot a }) 
                A({\bf Q,k}),
\label{eq:F_EQM}
\end{eqnarray}
with
\begin{eqnarray}
 A({\bf Q,q}) &=& S^-_{\bf Q/2-q} T^-_{\bf Q/2+q} |0> \\
 \gamma({\bf Q,q}) &=& \omega_s({\bf Q/2+q}) + \omega_t({\bf Q/2-q})  
\end{eqnarray}
where $d$ is the dimensionality of the system. 
The total momentum of the excitation is ${\bf Q}$ and the relative momentum 
of the combined spin and pseudo-spin excitation is ${\bf k}$. In order to check if there is a self-consistent solution
of the equation of motion, equation~(\ref{eq:F_EQM}) is iterated once. Then a set of equations
is found that can be represented in a matrix $|M_{\alpha,\beta}|$ of order $d$ for hyper-cubic lattices. 
In a three-dimensional system $\alpha,\beta = x,y,z$. 
The set of equations has a solution if, and only if,
\begin{eqnarray}
 Det \ | \delta_{\alpha,\beta} - M_{\alpha,\beta} | = 0
\label{eq:det}
\end{eqnarray}
with
\begin{eqnarray}
  M_{\alpha,\beta}&(&{\bf Q}) = -8 J_{st} \ (\frac{a}{2 \pi})^d \nonumber \\ 
                  &\cdot& \int d{\bf k}     
                   \frac{ (\cos Q_{\alpha}/2 - \cos k_{\alpha})(\cos Q_{\beta}/2 - \cos k_{\beta}) }
                           { \omega - \gamma({\bf Q,k}) }
\label{eq:bound}
\end{eqnarray}
If the system is one-dimensional the condition above reduces to
\begin{eqnarray}
  1 = -\frac{8 J_{st}}{2 \pi}  \int_{-\pi}^{\pi} d k
                      \frac{ (\cos Q/2 - \cos k)^2}
                           { \omega -\epsilon_s (k+Q/2) - \epsilon_t (k-Q/2) },
\label{eq:bound_1D}
\end{eqnarray}
where the lattice spacing is set to unity.
This expression resembles the condition for the existence of a two-magnon bound state
in a pure ferromagnetic Heisenberg model~\cite{Wortis63}.

\subsection{Bound states}

We restrict ourselves to the ferro-ferro phase, i.e. according to figure~\ref{fig:phases_2p}, $2J_{st}<J_s+J_t$
and $J_{st}>0$,
and determine the bound state energies from condition~(\ref{eq:det}) 
and~(\ref{eq:bound}). The integrals appearing
in the latter equation are, as in the two-magnon problem, by no means trivial and can only be determined
analytically in special cases. In general these equations can be solved numerically.\\
Before examining the integrals in detail, let us first turn to the excitations that lie in
the continuum.
The continuum of excitations starts at the energy $\omega_c({\bf Q})$ where the denominator of the 
matrix elements $M_{\alpha,\beta}$ starts to diverge. The condition for this is:
\begin{eqnarray}
\omega_c({\bf Q}) = {\rm Min}_{\bf k}[ \epsilon_s({\bf Q/2+k}) + \epsilon_t({\bf Q/2-k}) ]
\label{eq:cont}
\end{eqnarray}
The single (pseudo-)spin wave excitation energy {\it always}
lies in the two-particle continuum:
\begin{eqnarray}
\omega_s({\bf Q}) &\geq& \omega_c({\bf Q}) \nonumber \\
\omega_t({\bf Q}) &\geq& \omega_c({\bf Q}) \nonumber.
\end{eqnarray}
The explanation for this is that in a combined 
spin- and orbital-excitation the momentum is distributed over the two sub-systems and because the 
excitation spectrum in the ferro-ferro case is non-linear and concave ($[1-\cos]$-like), the 
energy of two excitations with smaller momenta can be lower than the energy of one excitation 
with large momentum. 

First we consider the {\em one-dimensional} case. The integral in equation~(\ref{eq:bound_1D}) 
can be reduced to the form
\begin{eqnarray}
\int \frac{(a - \cos k)^2}{1+ b \cos k + c \sin k} dk,
\end{eqnarray} 
which is known analytically. The condition that the integral be equal to unity is given
by the roots of a third order polynomial. Not so much is gained following
this procedure, as the resulting expressions are very long and complicated.
So let us consider some special momenta in the Brillouin zone.\\

${\bf Q=0}$\\
For $Q=0$ the condition~(\ref{eq:bound_1D}) for a bound state at
$\omega=0$ reduces to: $2J_{st}=J_s+J_t$, which is exactly at the phase boundary of the 
ferro-ferro phase in the two-site phase diagram, figure~\ref{fig:phases_2p}.  
A bound state at negative $\omega$ appears when $2J_{st}>J_s+J_t$. This
simply means that the ground state is not stable (by exciting it energy is gained), which 
can be expected as the two-site phase diagram shows that the antiferro-antiferro
phase is the ground state in this parameter range. The ferro-ferro ground state
is found to be exactly stable in the parameter ranges shown in figure~\ref{fig:phases_2p},
indicating that the two-site approximation gives a good prediction for the
ground state spin and orbital order, whereas the mean-field solution fails to
predict the right ordering.\\

\begin{figure}
      \epsfysize=50mm
      \centerline{\epsffile{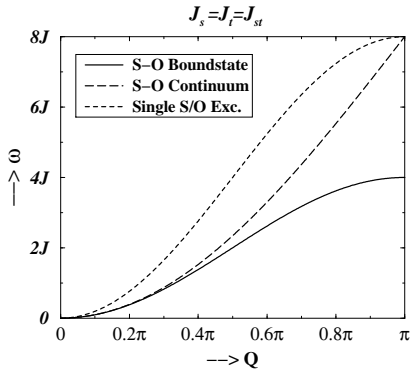}}
\caption{Dispersion of the spin-orbital bound state, spin-orbital continuum, single spin and single
orbital excitations in a one-dimensional ferro-ferro system where $J_s=J_t=J_{st}$. The single spin
and single orbital excitation energies coincide in this case. The unit of
energy is $J=J_{s}$.}
\label{fig:bound_sym}
\end{figure}

${\bf Q=\pi}$\\
By treating this special case, we can prove that there is {\it always} a bound state in the ferro-ferro
system, at least in one-dimension.
When $Q_{\alpha} = \pi$ for all $\alpha$, at the corner of the Brillouin zone, the equations simplify 
considerably. The off-diagonal matrix elements in equation~(\ref{eq:det}) 
vanish so that
$M_{\alpha,\beta}= D \delta_{\alpha,\beta}$. This yields
\begin{eqnarray}
\omega_b^{1d}({\pi}) = 2(J_s+J_t) -(J_s-J_t)^2/4J_{st}  \\
\omega_c^{1d}({\pi}) = 4( \ {\rm Min}[J_s,J_t] +J_{st} ),
\label{eq:pi_bound_1d}
\end{eqnarray}
From these equations follows that $\omega_b^{1d}({\pi}) < \omega_c^{1d}({\pi})$ for 
any $J_s$, $J_t$ and $J_{st}$, except when $J_{st}=|J_s-J_t|$, where 
$\omega_b^{1d}({\pi}) = \omega_c^{1d}({\pi})$.
This proves that at $Q=\pi$ there is always a bound state.\\

At $\omega({\bf Q}) \rightarrow \omega_c({\bf Q})$ in 1D\\
For energies approaching the continuum ($\omega({\bf Q}) \rightarrow \omega_c({\bf Q})$), 
the integrand in equation~(\ref{eq:bound_1D}),
diverges as $k^{-1}$ (except for $Q=0$), making the integral logarithmically
divergent. We can conclude from this that for any $Q$ (except for $Q=0$) there is always a point 
between $-\infty < \omega < \omega_c( Q)$ where the integral
is equal to unity. 

From the considerations above we can conclude that for the one-dimensional system
in the range $0 < {\bf Q} \leq \pi$ a spin-orbital bound state always exists and that this
bound state is the lowest lying elementary excitation of the ferro-ferro 
phase of 
model~(\ref{eq:HST}).
It is by definition lower in energy than the spin-orbital continuum,
which in turn is lower than the single (pseudo-)spin excitations.
Before illustrating the statement above with numerical examples, let us consider one more
special case, namely:\\

${\bf J_s=J_t=J_{st} }$ \\
For these parameters the system is exactly at the phase boundary of the ferro-ferro ground state.
Now equation~(\ref{eq:bound_1D}) takes a particularly simple form, and the solution for the bound state
for the one-dimensional system is
\begin{eqnarray}
\omega_b^{*}(Q) = 2J_{st} (1- \cos Q).
\end{eqnarray}
For the lower bound of the continuum, the single-spin and the single-orbital excitations one obtains 
in this case
\begin{eqnarray}
\omega_c^{*}(Q) = 8J_{st} (1- \cos Q/2), \\
\omega_s^{*}(Q) =\omega_t^{*}(Q) = 4J_{st} (1- \cos Q).
\end{eqnarray}
In figure~\ref{fig:bound_sym} the dispersion relations of these excitations are shown. 
The spin-orbital bound state is always the lowest
energy excitation of the system. Note that for small $Q$ the spin-orbital continuum and the bound state are
very close in energy and their energy difference is of the order of $\frac{J_{st}}{16} Q^4$.

\begin{figure}
      \epsfysize=45mm
      \centerline{\epsffile{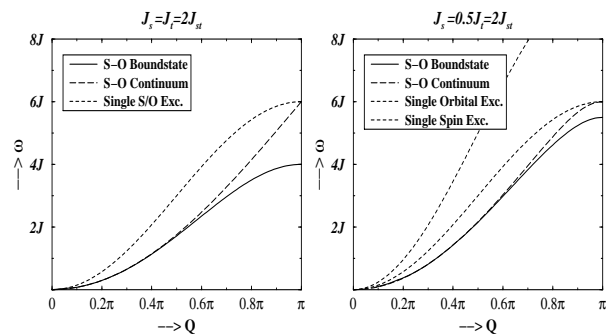}}
\caption{Dispersion of the spin-orbital bound state, spin-orbital continuum, single spin and single
orbital excitations in a one-dimensional ferro-ferro system. The unit of
energy is $J=J_{s}$.}
\label{fig:bound_asym}
\end{figure}

The numerical solution of the bound state equation for a one-dimensional system is shown in 
figure~\ref{fig:bound_asym} for two parameter sets. For small momenta the bound state is always
very close to the two-particle continuum. When $J_{st}$ is reduced, the single spin, single
orbital and continuum shift down in energy, approaching the bound state energy.
This can be expected since in the case when $J_{st}=0$ and $J_{s}=J_{t}$ the lower
bound of the continuum and the single spin and orbital excitation spectra 
all coincide.

In the right 
part of figure~\ref{fig:bound_asym} a case is shown where the single spin and single orbital
excitations have a different dispersion, i.e. $J_{s} \neq J_{t}$. The single orbital excitation
is shifted up in energy with respect to the single spin excitation. The bound state,  
continuum and magnon excitations are close in energy, and in the limit that $J_{t} \rightarrow \infty$
all three coincide, as can be expected.

\begin{figure}
      \epsfysize=50mm
      \centerline{\epsffile{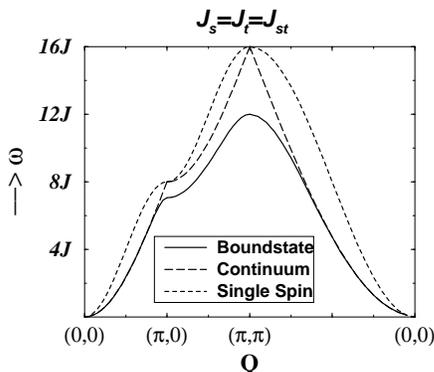}}
\caption{Dispersion of the spin-orbital bound state and spin-orbital continuum, single spin and single
orbital excitations
in a two dimensional ferro-ferro system. The unit of
energy is $J=J_{s}$.}
\label{fig:bound_2D_sym}
\end{figure}

In figure~\ref{fig:bound_2D_sym} a typical result for a two dimensional system is shown. It is
found numerically that there always exists at least one bound state, also in two dimensions.
Similar to the 1D case, the bound state is only well separated from the continuum at the edge
of the Brillouin-zone. If $J_s=J_t$ it can be shown that
\begin{eqnarray}
\omega_b(\pi,0) = 4(2J_s +2J_{st} - \sqrt{2J_{st}^2+ 2J_s J_{st} +J_s^2})
\end{eqnarray}
and
\begin{eqnarray}
\omega_b(\pi,\pi) = 4(2J_s +J_{st}),
\end{eqnarray}
so that at these points the bound state is considerably lower than the continuum.

\subsection{Almost Degeneracy}

In the sub-sections above we assumed that the two atomic levels are completely degenerate.
Crystal fields  generally split the two levels. Within the approach above it
is fairly easy to treat this energy splitting.
Suppose the energy difference between the orbitals $\alpha$ and $\beta$ is $\Delta$.
The Hamiltonian to be added to equation~(\ref{eq:ham_general}) is
\begin{eqnarray}
H_{\Delta} &=& \Delta/2 \sum_{i,\sigma} ( n_{i,\sigma,\beta} -n_{i,\sigma,\alpha}) \nonumber \\
           &=& \Delta \sum_{i} T^z_i,
\label{eq:orb_mag}
\end{eqnarray}
where the last equality follows from the definition of the T-operators. 
The level-splitting
manifests itself in the pseudo-spin language as a magnetic field for the orbitals.
Carrying through the calculation for the excitations leads to renormalized spin-waves,
orbital-waves and bound states:
\begin{eqnarray}
\bar{\omega}_s &=& {\omega}_s \nonumber \\
\bar{\omega}_t &=& {\omega}_t + \Delta \nonumber \\
\bar{\omega}_{st} &=& {\omega}_{st} + \Delta.
\end{eqnarray}

\begin{figure}
      \epsfysize=45mm
      \centerline{\epsffile{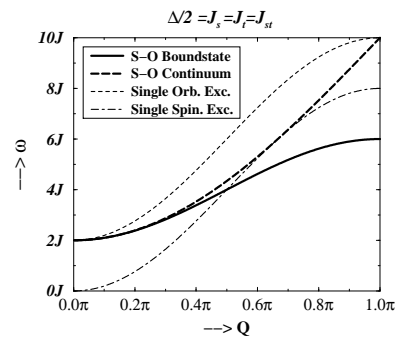}}
\caption{Dispersion of the spin-orbital bound state, spin-orbital continuum, single spin and single
orbital excitations for an almost degenerate 1D  ferro-ferro system where $\Delta/2=J_s=J_t=J_{st}$. 
The unit of energy is $J=J_{s}$.}
\label{fig:bound_sym_nondeg}
\end{figure}

The spin-wave spectrum is not affected, but the "magnetic field" for the orbitals causes a gap in
the orbital excitation spectrum. This can be expected since due to the magnetic field 
there is no breaking of a continuous symmetry in the orbital case, and hence no Goldstone mode.
This is reflected in the bound state energy being gapped.
The equalities above permit a convenient generalization of the results derived for the
fully degenerate system to the situation where the levels are non- (or almost-) degenerate,
as illustrated in figure~\ref{fig:bound_sym_nondeg}, where the dispersion of the bound state in the case of
an orbital energy splitting of $\Delta = 2J_{s}$ is shown.

From the results in this section one can also understand what will be the situation in the general
case discussed at the end of the introduction. Namely the orbital part of the effective spin pseudo-spin
Hamiltonian is in the general case anisotropic, containing both terms of the 
type~(\ref{eq:orb_mag})
and Ising-like terms $T^z_i T^z_j$. The situation then will be simular to the one
discussed above: there will appear a gap in the orbiton spectrum (see also~\cite{Ishihara96}). 
The spin-orbital bound state can still exist below the combined spin-orbital continuum (but in
general above pure spin waves) and one can show that the conditions for their existence
will be even less stringent than in the case of gapless orbitons. However, as can be seen
from figure~\ref{fig:bound_sym_nondeg}, these bound states will not be the lowest excitations
in the system, at least not in the whole Brillouin zone. Nevertheless they will lie lower  then the
orbital waves themselves, and they have definitely to be taken into account in a general treatment
of properties of such systems.
Note also that the orbital spectrum may still be gapless, even with realistic hopping integrals 
$t_{ij}^{\alpha \beta}$ due to the orbital-lattice symmetry; this can lead to the particular features of such
systems like quantum disorder~\cite{Feiner97}; the nature of the elementary excitations plays
a crucial role in this problem~\cite{Khal97}.

\section{Discussion}

We studied the low energy excitations of a two-fold degenerate Hubbard model
in the strong coupling limit with on average one electron per site and symmetric
hopping integrals.  We observe that besides the separate spin and orbital excitations,
combined excitations (spin-orbiton bound states) can exist and can even be the
lowest lying elementary excitation of the system.
In general overlap integrals explicitely depend on
the symmetry of the orbitals;  this, together with the Hund's rule coupling, may
break the continuous rotational symmetry of the orbital channel. 
In this case one can expect that the eventual spin-orbital bound states, which are gappless in the 
simplest case, become gapped and may be the lowest lying excitation only at larger momenta.

If not all interactions are ferromagnetic, there is a priori no reason to expect that the
orbital and spin-orbital bound states behave qualitatively different from the ferro-case,
although the quantum fluctuations might, or might not, destroy long range
order~\cite{Feiner97,Khal97,Feiner_prep}. 
This interesting aspect of the system with antiferromagnetic 
interactions still deserves further study.

The low energy collective modes of the orbitally degenerate Hubbard model certainly
contribute to the thermodynamic properties of the system, and should be observable
in for instance susceptibility and specific heat measurements. 
It should be stressed, however,  that the elementary excitations with predominantly
orbital character are in general gapped, and therefore lead only to moderate changes in thermodynamic
quantities. Experiments that are sensitive to higher energy scales might give direct evidence
for the existence of orbitons and spin-orbital bound states.
Orbital excitations, however, are strongly coupled to phonons and it might be difficult
to distinguish between these two contributions in experiment. One of the possibilities to pin them down
may be connected with the possible anomalies of the phonon dispersion relations induced by 
the orbital degrees of freedom.
Another possibility might be that, since the orbital
excitations also couple to the spin degrees of freedom, variations in the
spin order, for instance induced by an external magnetic field, will be reflected in the energy
and dispersion of the orbital related excitations and consequently in the properties of the
phonons mixed with orbitons. 

In conclusion, we studied a model Hamiltonian that captures the low energy behavior of a two-fold
degenerate Hubbard Hamiltonian. We presented the
phase diagram in the mean-field limit and in a two-site approach, revealing a rich
variety of phases where both the orbital and  spin degrees of freedom can be ordered (anti-)ferromagnetically.
We have shown that in this case there may exist, besides usual spin waves (magnons) also 
orbital waves (orbitons) and, most interestingly, the combined
spin-orbital excitation which can be visualized as bound states of magnons and orbitons.
In a fully degenerate system  the bound states
are found to be the lowest lying elementary excitations, both in one- and two-dimensions.
This shows that the elementary excitations
in orbitally degenerate strongly correlated electron systems in general may carry both spin 
and orbital character.

This work was financially supported by the Nederlandse Stichting voor Fundamenteel Onderzoek der Materie (FOM)
and the Stichting Scheikundig Onderzoek Nederland (SON), both financially supported by the Nederlandse
Organisatie voor Wetenschappelijk Onderzoek (NWO).


\end{document}